\documentclass[1opt,twocolumn,letter]{IEEEtran}
\usepackage{graphicx}
\DeclareGraphicsExtensions{.eps}
\graphicspath{{fig/}}
\usepackage[tight]{subfigure}
\usepackage{fancyhdr}
\usepackage{fancyvrb}
\usepackage{boxedminipage}
\usepackage{framed}
\usepackage{tikz}
\usepackage[sc, font=it]{caption}

\usepackage{times}

\usepackage[ruled,slide, vlined,linesnumbered]{algorithm2e}

\usepackage{multicol}

\newcommand{\ignore}[1]{{}}

\makeatletter \newcommand \listoftodos{\section*{Todo list} \@starttoc{tdo}}
\newcommand\l@todo[2]{
  \par\noindent \textit{#2}, \parbox{10cm}{#1}\par
} 
\makeatother

\newcommand{\Cross}{$\mathbin{\tikz [x=1.4ex,y=1.4ex,line width=.2ex, red] \draw (0,0) -- (1.5,1.5) (0,1.5) -- (1.5,0);}$}%

\usetikzlibrary{decorations.text}
\usetikzlibrary{decorations.pathmorphing}

\begin{document}

\title{Exploiting Redundant Computation\\
  in Communication-Avoiding Algorithms\\
  for Algorithm-Based Fault Tolerance}

\author{
\begin{tabular}{cc}
Camille Coti\\
\end{tabular}\\
\textit{LIPN, CNRS, UMR 7030} \\
\textit{Universit\'e Paris 13, Sorbonne Paris Cit\'e}\\
\textit{F-93430, Villetaneuse, France}\\
\textit{camille.coti@univ-paris13.fr}
}

\date{} 

\maketitle

\begin{abstract}
Communication-avoiding algorithms allow redundant computations to
minimize the number of inter-process communications. In this paper, we
propose to exploit this redundancy for fault-tolerance purpose. We
illustrate this idea with QR factorization of tall and skinny
matrices, and we evaluate the number of failures our algorithm can
tolerate under different semantics. 
\end{abstract}

\section{Introduction}
\label{sec:intro}

Faut tolerance for high performance distributed applications can be
achieved at system-level or application-level. System-level fault tolerance is transparent for the
application and requires a specific middleware that can restart the
failed processes and ensure coherent state of the application
\cite{fgcs08,BLKC04}. 

Application-level fault tolerance requires the application itself to
handle the failures and adapt to them. Of course, it implies that the
middleware that supports the distributed execution must be robust
enough to survive the failures and provide the application with
primitives to handle them \cite{FTMPI}. Moreover, it requires that
the application uses fault-tolerant algorithms that can deal with
process failures \cite{BDDL09}.

Recent efforts in the MPI-3 standardization process \cite{mpi3}
defined an interface for a mechanism called \emph{User-Level Failure
  Mitigation} (ULFM) \cite{BBHHBD13} and \emph{Run-Through
  Stabilization} \cite{HGBBPS11}.

This paper deals with the QR factorization of tall and skinny
matrices, and provide three fault-tolerant algorithms in the context
of ULFM. We give the robustness of each algorithm, the semantics of
the fault tolerance and we detail the behavior during failure-free
execution and upon failures. 
\section{Algorithm-Based Fault Tolerance}
\label{sec:abft}

FT-MPI~\cite{FTMPI, FGBACPLD04} defined four error-handling semantics
that can be defined on a communicator. \emph{SHRINK} consists in
reducing the size of the communicator in order to leave no hole in it
after a process of this communicator died. As a consequence, if one
process $p$ which is part of a communicator of size $N$ dies, after
the failure the communicator has $N-1$ processes numbered in
$[0,N-2]$. On the opposite, \emph{BLANK} leaves a hole in the
communicator: the rank of the dead process is considered as invalid
(communications return that the destination rank is invalid), and
surviving processes keep their original ranks in $[0,N-1]$. While
these two semantics survive failures with a reduced number of
processes, \emph{REBUILD} spawns a new process to replace the dead
one, giving it the place of the dead  process in the communicators it
was part of, including giving it the rank of the dead process. Last,
the \emph{ABORT} semantics corresponds to the usual behavior of
non-fault-tolerant applications: the surviving processes are
terminated and the application exits.

Using the first three semantics, programmers can integrate
 failure-recovery strategies directly as part of the algorithm that
 performs the computation. For instance, diskless
 checkpointing \cite{PLP98} uses the memory of other processes to save
 the state of each process. Arithmetic on the state of the processes
 can be used to store the checksum of a set of
 processes \cite{CFGLABD05}. When a process fails, its state can be
 recovered from the checkpoint and the states of the surviving
 processes. This approach is particularly interesting for iterative
 processes. Some matrix operations exhibit some properties on this
 checkpoint, such as \emph{checkpoint invariant} for LU
 factorization \cite{DBBHD12}. 

A proposal for \emph{run-through stabilization} introduced new
constructs to handle failures at communicator-level 
\cite{HGBBPS11}. Other mechanisms, at process-level, have been
integrated as a proposal in the MPI 3.1 standard draft \cite[ch
15]{mpi-3.1}. It is called \emph{user-level failure
mitigation} \cite{BBHHBD13}. Failures are detected when an operation
involving a failed process fails and returns an error. As a
consequence, operations that do not involve any failed process can
proceed unknowingly.  

\section{Fault-Tolerant, Communication-Avoiding Algorithms}
\label{sec:ftca}

Communication-avoiding algorithms were introduced on \cite{CAQR}. The
idea is to minimize the number of inter-process communications, should
it involve additional computations. These algorithms perform well on
current architectures, ranging from multicore
architectures \cite{DGG10} to aggregations of clusters \cite{ACDHL10},
because of the speed difference between computations and data
movements. 

As seen in the examples cited in section \ref{sec:abft}, tolerating
failures requires some form of \emph{redundancy}, such as checkpoints
or checksums stored in additional processes \cite{BDDL09}. 

In this paper, we propose to exploit the redundant computations made
by communication-avoiding algorithms for fault-tolerance purpose. In
this section we illustrate this idea with a communication-avoiding
algorithm for tall and skinny matrices (TSQR). This algorithm can be
used to compute the QR factorization of matrices with a lot of rows
and few columns, or as a panel factorization for QR
factorization \cite{HLAD09}. 

\subsection{Computing the R with TSQR}
\label{sec:ftca:tsqr}

The TSQR relies on successive steps that consist of local QR
factorizations, involving \emph{no inter-process communications}, and
one inter-process communication. Initially, the matrix is decomposed
in submatrices, each process holding one submatrix. On the first step,
each process performs a QR factorization on its local submatrix. Then
odd-numbered processes send the resulting $\widetilde{R}$ to the previous
even-numbered process: rank 1 sends to rank 0, rank 3 sends to rank
2\dots. The algorithm itself is given by Algorithm \ref{algo:tsqr}.

Even-numbered processes concatenate the two $\widetilde{R}$
matrices by creating a new matrix whose upper half is the
$\widetilde{R}$ it has computed and whose bottom half is the
$\widetilde{R}$ it has received. Then the odd-numbered process is done
with its participation to the computation of the $R$. If the $Q$
matrix is computed, it will work again when the moment comes, after
the computation of the $R$ is done.

\begin{figure}
  \resizebox{\linewidth}{!}{
\begin{tikzpicture}
\huge

\foreach \y in {0, 1, 2, 3}{
  \path[draw,thick] ( 0, -4*\y ) rectangle ( 1.8, -4*\y-3.8 );
    \node at ( -1, -4*\y-.8 ) {$\mathbf{P_\y}$ };
    \node at ( 1, -4*\y-.8 ) {$A_\y$ };
}


\foreach \y in {0, 1, 2, 3}{
  \path[draw,thick] ( 4, -4*\y ) rectangle ( 5.8, -4*\y-3.8 );
    \node at ( 5.2, -4*\y-.6 ) {$R_\y$ };
    \node at ( 4.5, -4*\y-2.5 ) {$V_\y$ };
    \draw [dashed, thick] ( 4, -4*\y ) -- ( 5.8, -4*\y-1.8 );
    \draw [ dashed, ->, thick ] (2, -4*\y-1) -- (3.8, -4*\y-1 );
}
  \node at ( 3, 1.5 ) {\textit{QR} };


  \foreach \y in {0, 2}{
  \path[draw,thick] ( 8, -4*\y ) rectangle ( 9.8, -4*\y-3.6 );
    \node at ( 9.2, -4*\y-.6 ) {$R_\y$ };
    \pgfmathsetmacro\z{\y + 1};
    \node at ( 9.2, -4*\y-2.4 ) {$R_{\pgfmathprintnumber{\z}}$ };
   \draw [dashed, thick] ( 8, -4*\y-1.8 ) -- ( 9.8, -4*\y-1.8 );
  \draw [dashed, thick] ( 8, -4*\y ) -- ( 9.8, -4*\y-1.8 );
    \draw [dashed, thick] ( 8, -4*\y-1.8 ) -- ( 9.8, -4*\y-3.6 );
     \draw [ dashed, ->, thick ] (6, -4*\y-5) -- (7.8, -4*\y-2.8 );
   \draw [ dashed, ->, thick ] (6, -4*\y-1) -- (7.8, -4*\y-1 );
}

  \node at ( 7, 1.5 ) {\textit{Send/Recv} };


  \foreach \y in {0, 2}{
  \path[draw,thick] ( 12, -4*\y ) rectangle ( 13.8, -4*\y-3.6 );
    \node at ( 13.2, -4*\y-.6 ) {$R_\y'$ };
    \node at ( 12.5, -4*\y-2.5 ) {${V_\y}'$ };
    \draw [dashed, thick] ( 12, -4*\y ) -- ( 13.8, -4*\y-1.8 );
  \draw [ dashed, ->, thick ] (10, -4*\y-1) -- (11.8, -4*\y-1 );

}
  \node at ( 11, 1.5 ) {\textit{QR} };
 

  \path[draw,thick] ( 16, 0 ) rectangle ( 17.8, -3.6 );
  \draw [dashed, thick] ( 16,0 ) -- ( 17.8, -1.8 );
    \draw [dashed, thick] ( 16, -1.8 ) -- ( 17.8, -3.6 );
    \draw [dashed, thick] ( 16, -1.8 ) -- ( 17.8, -1.8 );
    \node at ( 17.2, -.6 ) {$R_0'$ };
    \node at ( 17.2, -2.4 ) {$R_2'$ };
  \draw [ dashed, ->, thick ] (14, -1) -- (15.8, -1 );
  \draw [ dashed, ->, thick ] (14, -9) -- (15.8, -3 );

  \node at ( 15, 1.5 ) {\textit{Send/Recv} };


  \path[draw,thick] ( 20, 0 ) rectangle ( 21.8, -3.6 );
    \node at ( 21.2, -.6 ) {$R$ };
    \node at ( 20.5, -2.5 ) {$V$ };
  \draw [ dashed, ->, thick ] (18, -1) -- (19.8, -1 );
  \draw [dashed, thick] ( 20,0 ) -- ( 21.8, -1.8 );

  \node at ( 19, 1.5 ) {\textit{QR} };

\end{tikzpicture}
  }
\captionof{figure}{\label{fig:tsqr}Computing the R of a matrix using a
TSQR factorization on 4 processes.}
\end{figure}
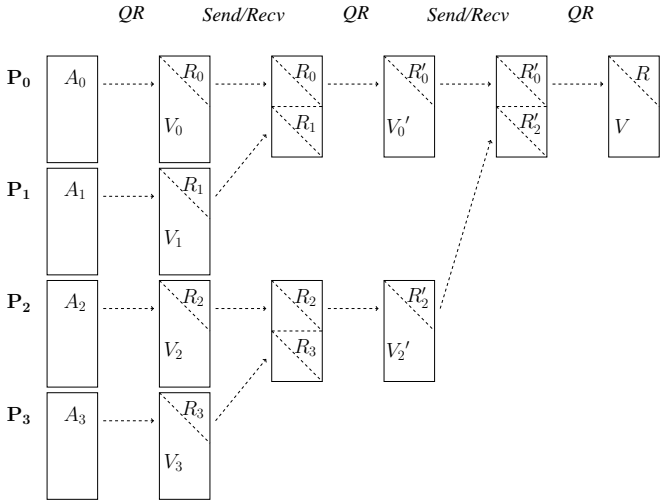

Even-numbered processes perform a local QR factorization of the
resulting matrix, and produce another $\widetilde{R}$. A similar
communication and concatenation step is performed between processes of
rank $r \pm 2^{step}$, if $r$ denotes the rank of each process. An
illustration of this communication, recombination and local
computation process on four processes is depicted by Figure \ref{fig:tsqr}.

At each step, half of the participating processes send their
$\widetilde{R}$ and are done. The other half receive an
$\widetilde{R}$, concatenate it with their own $\widetilde{R}$ and
perform a local QR factorization. This communication-computation
progression forms a binary reduction tree \cite{L10}. 

We can see on this example that once it has sent its $\widetilde{R}$,
each process becomes idle. Eventually, process 0  is the only working
process that remains. Half of the processes are idle after the first
step, one quarter are idle after the second step, and so on until only
one process is working at the end.

{\scriptsize
  \begin{algorithm}
    \SetKwFunction{isSender}{isSender}
    \SetKwFunction{isReceiver}{isReceiver}
    \SetKwFunction{mybuddy}{myBuddy}
    \SetKwFunction{send}{send}
    \SetKwFunction{recv}{recv}
    \SetKwFunction{done}{done}
    \SetKwFunction{qr}{QR}
    \SetKwFunction{concat}{concatenate}
    \KwData{Submatrix A}
    Q, R = \qr{ A }\;
    step = 0 \;
    
    \While{ ! \done{} }{
      \If{ \isSender{ step } } { \tcc{I am a sender}
        b = \mybuddy{ step } \;
        \send{ R, b } \;
        \Return \;
      }
      \Else { \tcc {I am a receiver}
        b = \mybuddy{ step } \;
        \recv{ R', b } \;
        A = \concat{ R, R' }\;
        Q, R = \qr{ A }\;
      }
      step++ \;
    }
    \tcc{The root of the tree reaches this point and owns the final $R$ }
    \Return R\; 
\caption{TSQR\label{algo:tsqr}} 
\end{algorithm}
}

\subsection{Redundant TSQR}
\label{sec:ftca:rtsqr}

We have seen in section \ref{sec:ftca:tsqr} that 1) only one process
eventually gets the resulting $R$ and 2) at each step, half of the
working processes get idle. The idea behind \emph{Redundant TSQR} is
to use these spare processes to produce copies of the intermediate
$\widetilde{R}$ factors, in order to tolerate process failures during
the intermediate steps. 

\subsubsection{Semantics}
\label{sec:ftca:rtsqr:sem}

With \emph{Redundant TSQR}, at the end of the computation, all the
processes get the final $R$ matrix. If some processes crash during the
computation but enough processes survive (see
section \ref{sec:ftca:rtsqr:rob}), the surviving processes have the
final $R$ factor. 

\subsubsection{Algorithm}
\label{sec:ftca:rtsqr:algo}

The basic idea is that when two processes communicate with each other,
instead of having one sender and one receiver that assembles the two
$\widetilde{R}$ matrices, the processes \emph{exchange} their
matrices. Both of them assemble the two matrices and both of them
proceed with the local QR factorization. This algorithm is given by
Algorithm \ref{algo:redtsqr}.

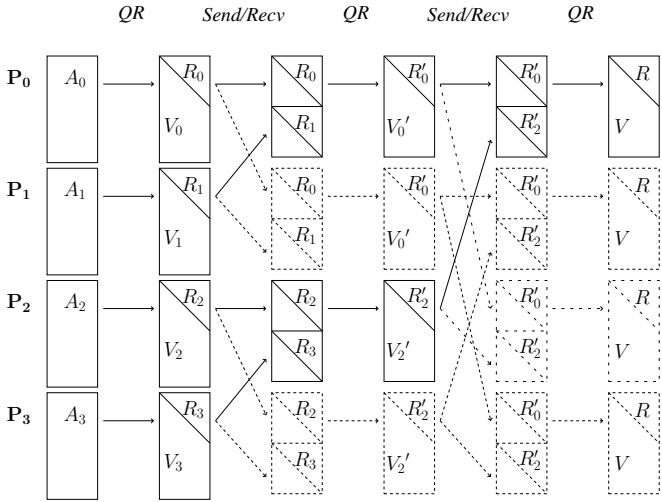
\begin{figure}
  \resizebox{\linewidth}{!}{
\begin{tikzpicture}
\huge


\foreach \y in {0, 1, 2, 3}{
  \path[draw,thick] ( 0, -4*\y ) rectangle ( 1.8, -4*\y-3.8 );
    \node at ( -1, -4*\y-.8 ) {$\mathbf{P_\y}$ };
    \node at ( 1, -4*\y-.8 ) {$A_\y$ };
}


\foreach \y in {0, 1, 2, 3}{
  \path[draw,thick] ( 4, -4*\y ) rectangle ( 5.8, -4*\y-3.8 );
    \node at ( 5.2, -4*\y-.6 ) {$R_\y$ };
    \node at ( 4.5, -4*\y-2.5 ) {$V_\y$ };
    \draw [ thick] ( 4, -4*\y ) -- ( 5.8, -4*\y-1.8 );
    \draw [ ->, thick ] (2, -4*\y-1) -- (3.8, -4*\y-1 );
}
  \node at ( 3, 1.5 ) {\textit{QR} };


  \foreach \y in {0, 2}{
  \path[draw,thick] ( 8, -4*\y ) rectangle ( 9.8, -4*\y-3.6 );
    \node at ( 9.2, -4*\y-.6 ) {$R_\y$ };
    \pgfmathsetmacro\z{\y + 1};
    \node at ( 9.2, -4*\y-2.4 ) {$R_{\pgfmathprintnumber{\z}}$ };
   \draw [thick] ( 8, -4*\y-1.8 ) -- ( 9.8, -4*\y-1.8 );
  \draw [ thick] ( 8, -4*\y ) -- ( 9.8, -4*\y-1.8 );
    \draw [thick] ( 8, -4*\y-1.8 ) -- ( 9.8, -4*\y-3.6 );
     \draw [  ->, thick ] (6, -4*\y-5) -- (7.8, -4*\y-2.8 );
   \draw [  ->, thick ] (6, -4*\y-1) -- (7.8, -4*\y-1 );
}


  \foreach \y in {1, 3}{
  \path[draw,thick, dashed] ( 8, -4*\y ) rectangle ( 9.8, -4*\y-3.6 );
    \pgfmathsetmacro\z{\y-1};
    \node at ( 9.2, -4*\y-.6 ) {$R_{\pgfmathprintnumber{\z}}$ };
    \pgfmathsetmacro\z{\y};
    \node at ( 9.2, -4*\y-2.4 ) {$R_{\pgfmathprintnumber{\z}}$ };
   \draw [thick, dashed] ( 8, -4*\y-1.8 ) -- ( 9.8, -4*\y-1.8 );
  \draw [ thick, dashed] ( 8, -4*\y ) -- ( 9.8, -4*\y-1.8 );
    \draw [thick, dashed] ( 8, -4*\y-1.8 ) -- ( 9.8, -4*\y-3.6 );
     \draw [  ->, thick, dashed ] (6, -4*\y+3) -- (7.8, -4*\y-.8 );
   \draw [  ->, thick, dashed ] (6, -4*\y-1) -- (7.8, -4*\y-3 );
}

  \node at ( 7, 1.5 ) {\textit{Send/Recv} };


  \foreach \y in {0, 2}{
  \path[draw,thick] ( 12, -4*\y ) rectangle ( 13.8, -4*\y-3.6 );
    \node at ( 13.2, -4*\y-.6 ) {$R_\y'$ };
    \node at ( 12.5, -4*\y-2.5 ) {${V_\y}'$ };
    \draw [ thick] ( 12, -4*\y ) -- ( 13.8, -4*\y-1.8 );
  \draw [  ->, thick ] (10, -4*\y-1) -- (11.8, -4*\y-1 );
}

  \foreach \y in {1, 3}{
  \path[draw,thick, dashed] ( 12, -4*\y ) rectangle ( 13.8, -4*\y-3.6 );
     \pgfmathsetmacro\z{\y-1};
   \node at ( 13.2, -4*\y-.6 ) {$R_{\pgfmathprintnumber{\z}}'$ };
    \node at ( 12.5, -4*\y-2.5 ) {${V_{\pgfmathprintnumber{\z}}}'$ };
    \draw [ thick, dashed] ( 12, -4*\y ) -- ( 13.8, -4*\y-1.8 );
  \draw [  ->, thick, dashed ] (10, -4*\y-1) -- (11.8, -4*\y-1 );
}

  \node at ( 11, 1.5 ) {\textit{QR} };
 

  \path[draw,thick] ( 16, 0 ) rectangle ( 17.8, -3.6 );
  \draw [ thick] ( 16,0 ) -- ( 17.8, -1.8 );
    \draw [ thick] ( 16, -1.8 ) -- ( 17.8, -3.6 );
    \draw [thick] ( 16, -1.8 ) -- ( 17.8, -1.8 );
    \node at ( 17.2, -.6 ) {$R_0'$ };
    \node at ( 17.2, -2.4 ) {$R_2'$ };
  \draw [  ->, thick ] (14, -1) -- (15.8, -1 );
  \draw [  ->, thick ] (14, -9) -- (15.8, -3 );


  \path[draw,thick,loosely dashed] ( 16, -8 ) rectangle ( 17.8, -11.6 );
  \draw [ thick, loosely dashed] ( 16,-8 ) -- ( 17.8, -9.8 );
    \draw [ thick, loosely dashed] ( 16, -9.8 ) -- ( 17.8, -11.6 );
    \draw [thick, loosely dashed] ( 16, -9.8 ) -- ( 17.8, -9.8 );
    \node at ( 17.2, -8.6 ) {$R_0'$ };
    \node at ( 17.2, -10.4 ) {$R_2'$ };
  \draw [  ->, thick,loosely  dashed ] (14, -1) -- (15.8, -9 );
 \draw [  ->, thick,loosely  dashed ] (14, -9) -- (15.8, -11 );


  \path[draw,thick,  dashed ] ( 16, -4 ) rectangle ( 17.8, -7.6 );
  \draw [ thick,  dashed] ( 16,-4 ) -- ( 17.8, -5.8 );
    \draw [ thick,  dashed] ( 16, -5.8 ) -- ( 17.8, -7.6 );
    \draw [thick,  dashed] ( 16, -5.8 ) -- ( 17.8, -5.8 );
    \node at ( 17.2, -4.6 ) {$R_0'$ };
    \node at ( 17.2, -6.4 ) {$R_2'$ };
  \draw [  ->, thick, dashed ] (14, -5) -- (15.8, -5 );
 \draw [  ->, thick, dashed ] (14, -13) -- (15.8, -7 );

  \path[draw,thick,  dashed ] ( 16, -12 ) rectangle ( 17.8, -15.6 );
  \draw [ thick,  dashed] ( 16,-12 ) -- ( 17.8, -13.8 );
    \draw [ thick,  dashed] ( 16, -13.8 ) -- ( 17.8, -15.6 );
    \draw [thick,  dashed] ( 16, -13.8 ) -- ( 17.8, -13.8 );
    \node at ( 17.2, -12.6 ) {$R_0'$ };
    \node at ( 17.2, -14.4 ) {$R_2'$ };
  \draw [  ->, thick, dashed ] (14, -5) -- (15.8, -13 );
 \draw [  ->, thick, dashed ] (14, -13) -- (15.8, -15 );

  \node at ( 15, 1.5 ) {\textit{Send/Recv} };


  \path[draw,thick] ( 20, 0 ) rectangle ( 21.8, -3.6 );
    \node at ( 21.2, -.6 ) {$R$ };
    \node at ( 20.5, -2.5 ) {$V$ };
  \draw [ ->, thick ] (18, -1) -- (19.8, -1 );
  \draw [ thick] ( 20,0 ) -- ( 21.8, -1.8 );

  \path[draw,thick, dashed] ( 20, -4 ) rectangle ( 21.8, -7.6 );
    \node at ( 21.2, -4.6 ) {$R$ };
    \node at ( 20.5, -6.5 ) {$V$ };
  \draw [ ->, thick,  dashed ] (18, -5) -- (19.8, -5 );
  \draw [ thick,  dashed] ( 20, -4 ) -- ( 21.8, -5.8 );

  \path[draw,thick, loosely dashed] ( 20, -8 ) rectangle ( 21.8, -11.6 );
    \node at ( 21.2, -8.6 ) {$R$ };
    \node at ( 20.5, -10.5 ) {$V$ };
  \draw [ ->, thick, loosely dashed ] (18, -9) -- (19.8, -9 );
  \draw [ thick, loosely dashed] ( 20, -8 ) -- ( 21.8, -9.8 );

  \path[draw,thick, dashed] ( 20, -12 ) rectangle ( 21.8, -15.6 );
    \node at ( 21.2, -12.6 ) {$R$ };
    \node at ( 20.5, -14.5 ) {$V$ };
  \draw [ ->, thick, dashed ] (18, -13) -- (19.8, -13 );
  \draw [ thick, dashed] ( 20, -12 ) -- ( 21.8, -13.8 );

  \node at ( 19, 1.5 ) {\textit{QR} };
\end{tikzpicture}
  }
\captionof{figure}{\label{fig:rtsqr}Computing the R of a matrix using
a TSQR factorization on 4 processes with redundant $\widetilde{R}$ factors.}
\end{figure}

This algorithm is represented on four processes in
Figure \ref{fig:rtsqr}. Plain lines represent the regular TSQR
pattern. During the first communication stage, the redundancies are
represented by dashed lines: $P_1$ and $P_3$ exchange data
with $P_0$ and $P_2$ respectively, and therefore obtain the same
intermediate matrices. Then same data exchange is performed during the
following step, resulting in a first level of redundancy (obtained
from the $P_0 \leftrightarrow P_2$ exchange), represented by loosely
dashed lines, and a secondary level of redundancy (obtained
from the $P_1 \leftrightarrow P_3$ exchange), represented by dashed
lines. 

{\scriptsize
  \begin{algorithm}
    \SetKwFunction{mybuddy}{myBuddy}
    \SetKwFunction{sendrecv}{sendrecv}
    \SetKwFunction{done}{done}
    \SetKwFunction{qr}{QR}
    \SetKwFunction{concat}{concatenate}
    \KwData{Submatrix A}
    Q, R = \qr{ A }\;
    step = 0 \;
    
    \While{ ! \done{} }{
      b = \mybuddy{ step } \;
      f = \sendrecv{ R, R', b } \;
      \If{ $FAIL$ == $f$ } {
        \Return \; \label{algo:redtsqr:failure}
      }
      A = \concat{ R, R' }\;
      Q, R = \qr{ A }\;
      step++ \;
    }
    \tcc{All the surviving processes reach this point and own the final $R$ }
    \Return R\; 
\caption{Redundant TSQR\label{algo:redtsqr}} 
\end{algorithm}
}

\subsubsection{Robustness}
\label{sec:ftca:rtsqr:rob}

We can see that at each step, the data exchange creates one extra copy
of each intermediate matrix. As a consequence, the redundancy rate
doubles at each step of the algorithm. Therefore, if $s$ denotes the
step number, the number of existing copies in the system is
$2^s$. Hence, this algorithm can tolerate $2^s - 1$ process failures. 

We can see that the number of failures that this algorithm can
tolerate increases as the computation progresses. This fact is a
direct consequence from the fact that the number of redundant copies
of the data is multiplied by 2 at each step. For instance, the
computation can proceed if no more than 1 process have failed by the
end of step 1, no more than 3 processes have failed by the end of step
2, etc. In the meantime, the number of failures in the system increase
with time: the longer a computation lasts, the more processes will
fail \cite{Reed2006293}. Therefore, the robustness of this algorithm
increases with time, which is consistent with the need for robustness. 

\subsubsection{Behavior upon failures}
\label{sec:ftca:rtsqr:failure}

When a process fails, the other processes proceed with the
execution. Processes that require data from the failed process end
their execution, and those that require data from ended processes end
theirs as well (see line \ref{algo:redtsqr:failure} of Algorithm
\ref{algo:redtsqr}). 

\begin{figure}
  \resizebox{\linewidth}{!}{
\begin{tikzpicture}
\huge


\foreach \y in {0, 1, 2, 3}{
  \path[draw,thick] ( 0, -4*\y ) rectangle ( 1.8, -4*\y-3.8 );
    \node at ( -1, -4*\y-.8 ) {$\mathbf{P_\y}$ };
    \node at ( 1, -4*\y-.8 ) {$A_\y$ };
}


\foreach \y in {0, 1, 2, 3}{
  \path[draw,thick] ( 4, -4*\y ) rectangle ( 5.8, -4*\y-3.8 );
    \node at ( 5.2, -4*\y-.6 ) {$R_\y$ };
    \node at ( 4.5, -4*\y-2.5 ) {$V_\y$ };
    \draw [ thick] ( 4, -4*\y ) -- ( 5.8, -4*\y-1.8 );
    \draw [ ->, thick ] (2, -4*\y-1) -- (3.8, -4*\y-1 );
}
  \node at ( 3, 1.5 ) {\textit{QR} };


  \foreach \y in {0, 2}{
  \path[draw,thick] ( 8, -4*\y ) rectangle ( 9.8, -4*\y-3.6 );
    \node at ( 9.2, -4*\y-.6 ) {$R_\y$ };
    \pgfmathsetmacro\z{\y + 1};
    \node at ( 9.2, -4*\y-2.4 ) {$R_{\pgfmathprintnumber{\z}}$ };
   \draw [thick] ( 8, -4*\y-1.8 ) -- ( 9.8, -4*\y-1.8 );
  \draw [ thick] ( 8, -4*\y ) -- ( 9.8, -4*\y-1.8 );
    \draw [thick] ( 8, -4*\y-1.8 ) -- ( 9.8, -4*\y-3.6 );
     \draw [  ->, thick ] (6, -4*\y-5) -- (7.8, -4*\y-2.8 );
   \draw [  ->, thick ] (6, -4*\y-1) -- (7.8, -4*\y-1 );
}


  \foreach \y in {1, 3}{
  \path[draw,thick, dashed] ( 8, -4*\y ) rectangle ( 9.8, -4*\y-3.6 );
    \pgfmathsetmacro\z{\y-1};
    \node at ( 9.2, -4*\y-.6 ) {$R_{\pgfmathprintnumber{\z}}$ };
    \pgfmathsetmacro\z{\y};
    \node at ( 9.2, -4*\y-2.4 ) {$R_{\pgfmathprintnumber{\z}}$ };
   \draw [thick, dashed] ( 8, -4*\y-1.8 ) -- ( 9.8, -4*\y-1.8 );
  \draw [ thick, dashed] ( 8, -4*\y ) -- ( 9.8, -4*\y-1.8 );
    \draw [thick, dashed] ( 8, -4*\y-1.8 ) -- ( 9.8, -4*\y-3.6 );
     \draw [  ->, thick, dashed ] (6, -4*\y+3) -- (7.8, -4*\y-.8 );
   \draw [  ->, thick, dashed ] (6, -4*\y-1) -- (7.8, -4*\y-3 );
}

  \node at ( 7, 1.5 ) {\textit{Send/Recv} };


  \path[draw,thick] ( 12, 0 ) rectangle ( 13.8, -3.6 );
    \node at ( 13.2, -.6 ) {$R_0'$ };
    \node at ( 12.5, -2.5 ) {${V_0}'$ };
    \draw [ thick] ( 12, 0 ) -- ( 13.8, -1.8 );
  \draw [  ->, thick ] (10, -1) -- (11.8, -1 );

  \foreach \y in {1, 3}{
  \path[draw,thick, dashed] ( 12, -4*\y ) rectangle ( 13.8, -4*\y-3.6 );
     \pgfmathsetmacro\z{\y-1};
   \node at ( 13.2, -4*\y-.6 ) {$R_{\pgfmathprintnumber{\z}}'$ };
    \node at ( 12.5, -4*\y-2.5 ) {${V_{\pgfmathprintnumber{\z}}}'$ };
    \draw [ thick, dashed] ( 12, -4*\y ) -- ( 13.8, -4*\y-1.8 );
  \draw [  ->, thick, dashed ] (10, -4*\y-1) -- (11.8, -4*\y-1 );
}


\node at ( 12.9, -9.2 )  {\Cross};
\node at ( 12.9, -10.3 )  {CRASH};

 \node at ( 11, 1.5 ) {\textit{QR} };
 

 \node at ( 16.9, -1.3 )  {STOP};



  \path[draw,thick,  dashed ] ( 16, -4 ) rectangle ( 17.8, -7.6 );
  \draw [ thick,  dashed] ( 16,-4 ) -- ( 17.8, -5.8 );
    \draw [ thick,  dashed] ( 16, -5.8 ) -- ( 17.8, -7.6 );
    \draw [thick,  dashed] ( 16, -5.8 ) -- ( 17.8, -5.8 );
    \node at ( 17.2, -4.6 ) {$R_0'$ };
    \node at ( 17.2, -6.4 ) {$R_2'$ };
  \draw [  ->, thick, dashed ] (14, -5) -- (15.8, -5 );
 \draw [  ->, thick, dashed ] (14, -13) -- (15.8, -7 );

  \path[draw,thick,  dashed ] ( 16, -12 ) rectangle ( 17.8, -15.6 );
  \draw [ thick,  dashed] ( 16,-12 ) -- ( 17.8, -13.8 );
    \draw [ thick,  dashed] ( 16, -13.8 ) -- ( 17.8, -15.6 );
    \draw [thick,  dashed] ( 16, -13.8 ) -- ( 17.8, -13.8 );
    \node at ( 17.2, -12.6 ) {$R_0'$ };
    \node at ( 17.2, -14.4 ) {$R_2'$ };
  \draw [  ->, thick, dashed ] (14, -5) -- (15.8, -13 );
 \draw [  ->, thick, dashed ] (14, -13) -- (15.8, -15 );

  \node at ( 15, 1.5 ) {\textit{Send/Recv} };


  \path[draw,thick, dashed] ( 20, -4 ) rectangle ( 21.8, -7.6 );
    \node at ( 21.2, -4.6 ) {$R$ };
    \node at ( 20.5, -6.5 ) {$V$ };
  \draw [ ->, thick,  dashed ] (18, -5) -- (19.8, -5 );
  \draw [ thick,  dashed] ( 20, -4 ) -- ( 21.8, -5.8 );

  \path[draw,thick, dashed] ( 20, -12 ) rectangle ( 21.8, -15.6 );
    \node at ( 21.2, -12.6 ) {$R$ };
    \node at ( 20.5, -14.5 ) {$V$ };
  \draw [ ->, thick, dashed ] (18, -13) -- (19.8, -13 );
  \draw [ thick, dashed] ( 20, -12 ) -- ( 21.8, -13.8 );

  \node at ( 19, 1.5 ) {\textit{QR} };
\end{tikzpicture}
  }
\captionof{figure}{\label{fig:rtsqr:crash}Redundant TSQR on 4
  processes with one process failure.}
\end{figure}
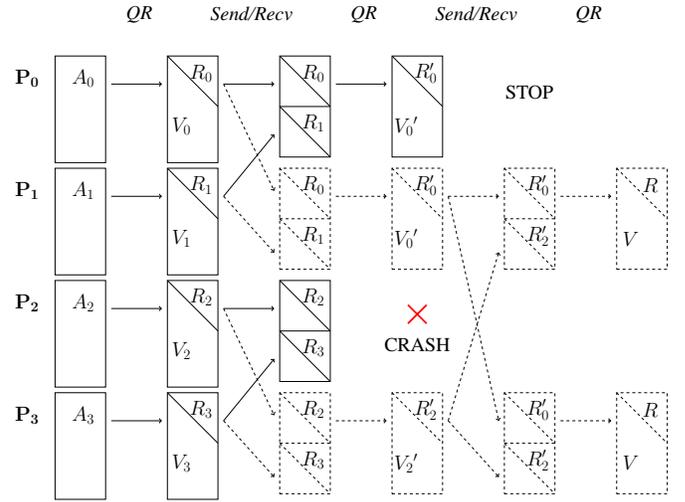

For instance, Figure \ref{fig:rtsqr:crash} represents the execution of
Redundant TSQR on four processes. Process $P_2$ crashes at the end of
the first step. The data it contained is also held by process $P_3$,
therefore the execution can proceed. However, process $P_0$ needs data
prom the failed process at the following step. Therefore, process
$P_0$ ends its execution. As a consequence, $P_0$ ends its
execution. At the end of the computation, the final result $R$ has
been computed by processes $P_1$ and $P_3$ and therefore, the final
result is available in spite of the failure. 

\subsection{Replace TSQR}
\label{sec:ftca:rptsqr}

\subsubsection{Semantics}
\label{sec:ftca:rptsqr:sem}

The semantics of \emph{Replace TSQR},is similar to the one with
\emph{Redundant TSQR}: at the end of the computation, all the
processes get the final $R$ matrix. If some processes crash during the
computation but enough processes survive (see section
\ref{sec:ftca:rptsqr:rob}), the surviving processes have the final $R$
factor.

\subsubsection{Algorithm}
\label{sec:ftca:rptsqr:algo}

The fault-free execution of the \emph{Replace TSQR} algorithm is
exactly the same as with \emph{Redundant TSQR} (see section
\ref{sec:ftca:rtsqr:algo}). The data held by processes along the
reduction tree of the initial TSQR algorithm is replicated on spare
processes that would normally stop their execution. 

The difference comes when a failure occurs. In this case, the process
that needs to communicate with another process gets an error when it
tries to communicate with it. Then, it finds a \emph{replica} of the
process it is trying to communicate with (line
\ref{algo:reptsqr:replica} of Algorithm \ref{algo:reptsqr}) and
exchanges its matrix with it. If no replica can be found alive, it
means that too many processes have failed and no extra copy of this
submatrix exist. Then the process exits. The algorithm is
described by Algorithm \ref{algo:reptsqr}.

{\scriptsize
  \begin{algorithm}
    \SetKwFunction{mybuddy}{myBuddy}
    \SetKwFunction{sendrecv}{sendrecv}
    \SetKwFunction{done}{done}
    \SetKwFunction{findreplica}{findReplica}
    \SetKwFunction{qr}{QR}
    \SetKwFunction{concat}{concatenate}
    \KwData{Submatrix A}
    Q, R = \qr{ A }\;
    step = 0;
    \While{ ! \done{} }{
      b = \mybuddy{ step } \;
      f = \sendrecv{ R, R', b } \;
      \While{ $FAIL$ == $f$ } {
        b = \findreplica{ b } \label{algo:reptsqr:replica} \;
        \If{ $None$ == $b$ } {
          \Return \; \label{algo:reptsqr:exit}
        }
        f = \sendrecv{ R, R', b } \;
      }
      A = \concat{ R, R' }\;
      Q, R = \qr{ A }\;
      step++ \;
    }
    \tcc{All the surviving processes reach this point and own the
      final $R$ } 
    \Return R\; 
\caption{Replace TSQR\label{algo:reptsqr}} 
\end{algorithm}
}

\subsubsection{Robustness}
\label{sec:ftca:rptsqr:rob}

We have seen in section \ref{sec:ftca:rptsqr:algo} that this algorithm
can keen progressing as long as there exists at least one copy of each
submatrix. We have seen in section \ref{sec:ftca:rtsqr:rob} that at
each step $s$, the number of existing copies in the system is
$2^s$. Hence, this algorithm can tolerate $2^s - 1$ process failures,
just like the \emph{Redundant TSQR} algorithm (see section
\ref{sec:ftca:rtsqr:rob}).

The difference between the \emph{Redundant TSQR} and the \emph{Replace
  TSQR} is that with the former, the processes that need to
communicate with a failed process exit, whereas with the latter, they
try to find a replica. Therefore, if the root of the tree does not
die, it holds the final result $R$ at the end of the computation.

\subsubsection{Behavior upon failures}
\label{sec:ftca:rptsqr:failure}

If a process fails, the processes that try to communicate with it fail
to do so and try to find a replica to communicate with. 

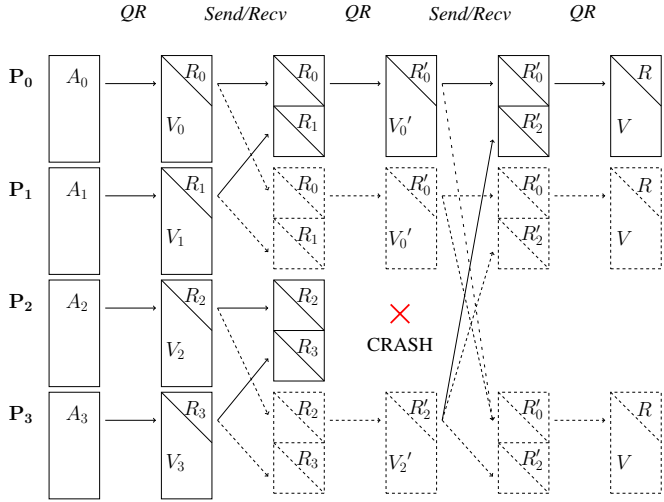
\begin{figure}
  \resizebox{\linewidth}{!}{
\begin{tikzpicture}
\huge


\foreach \y in {0, 1, 2, 3}{
  \path[draw,thick] ( 0, -4*\y ) rectangle ( 1.8, -4*\y-3.8 );
    \node at ( -1, -4*\y-.8 ) {$\mathbf{P_\y}$ };
    \node at ( 1, -4*\y-.8 ) {$A_\y$ };
}


\foreach \y in {0, 1, 2, 3}{
  \path[draw,thick] ( 4, -4*\y ) rectangle ( 5.8, -4*\y-3.8 );
    \node at ( 5.2, -4*\y-.6 ) {$R_\y$ };
    \node at ( 4.5, -4*\y-2.5 ) {$V_\y$ };
    \draw [ thick] ( 4, -4*\y ) -- ( 5.8, -4*\y-1.8 );
    \draw [ ->, thick ] (2, -4*\y-1) -- (3.8, -4*\y-1 );
}
  \node at ( 3, 1.5 ) {\textit{QR} };


  \foreach \y in {0, 2}{
  \path[draw,thick] ( 8, -4*\y ) rectangle ( 9.8, -4*\y-3.6 );
    \node at ( 9.2, -4*\y-.6 ) {$R_\y$ };
    \pgfmathsetmacro\z{\y + 1};
    \node at ( 9.2, -4*\y-2.4 ) {$R_{\pgfmathprintnumber{\z}}$ };
   \draw [thick] ( 8, -4*\y-1.8 ) -- ( 9.8, -4*\y-1.8 );
  \draw [ thick] ( 8, -4*\y ) -- ( 9.8, -4*\y-1.8 );
    \draw [thick] ( 8, -4*\y-1.8 ) -- ( 9.8, -4*\y-3.6 );
     \draw [  ->, thick ] (6, -4*\y-5) -- (7.8, -4*\y-2.8 );
   \draw [  ->, thick ] (6, -4*\y-1) -- (7.8, -4*\y-1 );
}


  \foreach \y in {1, 3}{
  \path[draw,thick, dashed] ( 8, -4*\y ) rectangle ( 9.8, -4*\y-3.6 );
    \pgfmathsetmacro\z{\y-1};
    \node at ( 9.2, -4*\y-.6 ) {$R_{\pgfmathprintnumber{\z}}$ };
    \pgfmathsetmacro\z{\y};
    \node at ( 9.2, -4*\y-2.4 ) {$R_{\pgfmathprintnumber{\z}}$ };
   \draw [thick, dashed] ( 8, -4*\y-1.8 ) -- ( 9.8, -4*\y-1.8 );
  \draw [ thick, dashed] ( 8, -4*\y ) -- ( 9.8, -4*\y-1.8 );
    \draw [thick, dashed] ( 8, -4*\y-1.8 ) -- ( 9.8, -4*\y-3.6 );
     \draw [  ->, thick, dashed ] (6, -4*\y+3) -- (7.8, -4*\y-.8 );
   \draw [  ->, thick, dashed ] (6, -4*\y-1) -- (7.8, -4*\y-3 );
}

  \node at ( 7, 1.5 ) {\textit{Send/Recv} };


  \foreach \y in {0}{
  \path[draw,thick] ( 12, -4*\y ) rectangle ( 13.8, -4*\y-3.6 );
    \node at ( 13.2, -4*\y-.6 ) {$R_\y'$ };
    \node at ( 12.5, -4*\y-2.5 ) {${V_\y}'$ };
    \draw [ thick] ( 12, -4*\y ) -- ( 13.8, -4*\y-1.8 );
  \draw [  ->, thick ] (10, -4*\y-1) -- (11.8, -4*\y-1 );
}

  \foreach \y in {1, 3}{
  \path[draw,thick, dashed] ( 12, -4*\y ) rectangle ( 13.8, -4*\y-3.6 );
     \pgfmathsetmacro\z{\y-1};
   \node at ( 13.2, -4*\y-.6 ) {$R_{\pgfmathprintnumber{\z}}'$ };
    \node at ( 12.5, -4*\y-2.5 ) {${V_{\pgfmathprintnumber{\z}}}'$ };
    \draw [ thick, dashed] ( 12, -4*\y ) -- ( 13.8, -4*\y-1.8 );
  \draw [  ->, thick, dashed ] (10, -4*\y-1) -- (11.8, -4*\y-1 );
}

  \node at ( 11, 1.5 ) {\textit{QR} };
 
  
    \node at ( 12.5, -9.2 )  {\Cross};
  \node at ( 12.5, -10.3 )  {CRASH};


  \path[draw,thick] ( 16, 0 ) rectangle ( 17.8, -3.6 );
  \draw [ thick] ( 16,0 ) -- ( 17.8, -1.8 );
    \draw [ thick] ( 16, -1.8 ) -- ( 17.8, -3.6 );
    \draw [thick] ( 16, -1.8 ) -- ( 17.8, -1.8 );
    \node at ( 17.2, -.6 ) {$R_0'$ };
    \node at ( 17.2, -2.4 ) {$R_2'$ };
  \draw [  ->, thick ] (14, -1) -- (15.8, -1 );
  \draw [  ->, thick ] (14, -13) -- (15.8, -3 );


  \draw [  ->, thick,loosely  dashed ] (14, -1) -- (15.8, -13 );


  \path[draw,thick,  dashed ] ( 16, -4 ) rectangle ( 17.8, -7.6 );
  \draw [ thick,  dashed] ( 16,-4 ) -- ( 17.8, -5.8 );
    \draw [ thick,  dashed] ( 16, -5.8 ) -- ( 17.8, -7.6 );
    \draw [thick,  dashed] ( 16, -5.8 ) -- ( 17.8, -5.8 );
    \node at ( 17.2, -4.6 ) {$R_0'$ };
    \node at ( 17.2, -6.4 ) {$R_2'$ };
  \draw [  ->, thick, dashed ] (14, -5) -- (15.8, -5 );
 \draw [  ->, thick, dashed ] (14, -13) -- (15.8, -7 );

  \path[draw,thick,  dashed ] ( 16, -12 ) rectangle ( 17.8, -15.6 );
  \draw [ thick,  dashed] ( 16,-12 ) -- ( 17.8, -13.8 );
    \draw [ thick,  dashed] ( 16, -13.8 ) -- ( 17.8, -15.6 );
    \draw [thick,  dashed] ( 16, -13.8 ) -- ( 17.8, -13.8 );
    \node at ( 17.2, -12.6 ) {$R_0'$ };
    \node at ( 17.2, -14.4 ) {$R_2'$ };
 \draw [  ->, thick, dashed ] (14, -13) -- (15.8, -15 );
  \draw [  ->, thick, dashed ] (14, -5) -- (15.8, -13 );

  \node at ( 15, 1.5 ) {\textit{Send/Recv} };


  \path[draw,thick] ( 20, 0 ) rectangle ( 21.8, -3.6 );
    \node at ( 21.2, -.6 ) {$R$ };
    \node at ( 20.5, -2.5 ) {$V$ };
  \draw [ ->, thick ] (18, -1) -- (19.8, -1 );
  \draw [ thick] ( 20,0 ) -- ( 21.8, -1.8 );

  \path[draw,thick, dashed] ( 20, -4 ) rectangle ( 21.8, -7.6 );
    \node at ( 21.2, -4.6 ) {$R$ };
    \node at ( 20.5, -6.5 ) {$V$ };
  \draw [ ->, thick,  dashed ] (18, -5) -- (19.8, -5 );
  \draw [ thick,  dashed] ( 20, -4 ) -- ( 21.8, -5.8 );

  \path[draw,thick, dashed] ( 20, -12 ) rectangle ( 21.8, -15.6 );
    \node at ( 21.2, -12.6 ) {$R$ };
    \node at ( 20.5, -14.5 ) {$V$ };
  \draw [ ->, thick, dashed ] (18, -13) -- (19.8, -13 );
  \draw [ thick, dashed] ( 20, -12 ) -- ( 21.8, -13.8 );

  \node at ( 19, 1.5 ) {\textit{QR} };
\end{tikzpicture}
  }
\captionof{figure}{\label{fig:rptsqr:crash}Replace TSQR on 4
  processes with one process failure.}
\end{figure}

For instance, Figure \ref{fig:rptsqr:crash} represents the execution of
Redundant TSQR on four processes. Process $P_2$ crashes at the end of
the first step. The data it contained is also held by process $P_3$,
therefore the execution can proceed. However, process $P_0$ needs data
prom the failed process at the following step. Therefore, process
$P_0$ ends its execution. As a consequence, $P_0$ fails to communicate
with $P_0$ and finds out that $P_3$ holds the same data as $P_2$. Then
$P_0$ exchanges data with $P_3$.

\subsection{Self-Healing TSQR}
\label{sec:ftca:shtsqr}

The previous algorithms described here, \emph{Redundant TSQR} (section
\ref{sec:ftca:rtsqr}) and \emph{Replace TSQR} (section
\ref{sec:ftca:rptsqr}) proceed with the execution without the dead
processes. Here we are describing an algorithm that replaces the dead
process with a new one.

\subsubsection{Semantics}
\label{sec:ftca:shtsqr:sem}

With \emph{Self-Healing TSQR}, at the end of the computation, all the
processes get the final $R$ matrix. If some processes crash during the
computation but enough processes survive at each step (see
section \ref{sec:ftca:shtsqr:rob}), the final number of processes is
the same as the initial number and all the processes have the
final $R$ factor. 

\subsubsection{Algorithm}
\label{sec:ftca:shtsqralgo}

This algorithm follows the same basic idea as \emph{Redundant TSQR}
(see section \ref{sec:ftca:rtsqr}) in a sense that at each step of the
computation, all the processes send or receive their $\widetilde{R}$
matrices and compute the $R$ of the resulting matrix. As a
consequence, the data required by the computation (the intermediate
submatrices represented in Figure \ref{fig:tsqr}) are
\emph{replicated}. This part is described by Algorithm
\ref{algo:shtsqr:comp} with the initialization described by Algorithm
\ref{algo:shtsqr:init}.

In this algorithm, the failed processes are replaced by newly spawned
ones. We have seen that the data contained by the failed process has
been replicated by the redundant computations. As a consequence, a
failed process can be recovered completely and a newly spawned process
can replace it: see Algorithm \ref{algo:shtsqr:restart}.

The fault-free execution of this algorithm is similar with the
execution represented by Figure \ref{fig:rtsqr}.

{\scriptsize
  \begin{algorithm}
    \SetKwFunction{mybuddy}{myBuddy}
    \SetKwFunction{sendrecv}{sendrecv}
    \SetKwFunction{done}{done}
    \SetKwFunction{shtsqr}{QR}
    \SetKwFunction{qr}{QR}
    \SetKwFunction{spawn}{spawnNew}
    \SetKwFunction{concat}{concatenate}
    \KwData{Submatrix A}
    Q, R = \qr{ A }\;
    step = 0 \;

    R = \shtsqr{ R, step} \; 
    
    \Return R\; 
\caption{Self-Healing TSQR: initialization\label{algo:shtsqr:init}} 
\end{algorithm}
}

{\scriptsize
  \begin{algorithm}
    \SetKwFunction{mytwin}{mytwin}
    \SetKwFunction{recv}{recv}
    \tcc{Gets my data from a process that holds the same as me.}
    t = \mytwin{} \;
    R, step = \recv{ t } \;
    \tcc{Proceed with the computation } 
    R = \shtsqr{ R, step } \;
    \tcc{At the end of the computation, this process holds the final $R$.}
    \Return R\; 
\caption{Self-Healing TSQR: process restart\label{algo:shtsqr:restart}} 
\end{algorithm}
}

{\scriptsize
  \begin{algorithm}
    \SetKwBlock{Function}{Function shtsqr( A, step ):}{end}
    \SetKwFunction{mytwin}{mytwin}
    \SetKwFunction{recv}{recv}
    \Function{ 
    Q, R = \qr{ A }\;
    
    \While{ ! \done{} }{
      b = \mybuddy{ step } \;
      f = \sendrecv{ R, R', b } \;
      \If{ $FAIL$ == $f$ } {\label{algo:shtsqr:failure}
        \spawn{ b } \;
      }
      A = \concat{ R, R' }\;
      Q, R = \qr{ A }\;
      step++ \;
    }
    \tcc{All the processes reach this point and own the final $R$ }
    \Return R\;
    }
    \caption{Self-Healing TSQR: computation\label{algo:shtsqr:comp}} 
  \end{algorithm}
}

\subsubsection{Robustness}
\label{sec:ftca:shtsqr:rob}

We have seen in \ref{sec:ftca:shtsqralgo} and
\ref{sec:ftca:rtsqr:rob} that at each step $s$, the data necessary for
each process from the original algorithm is replicated $2^s$ times on
other processes. As a consequence, this algorithm can tolerate $2^s -
1$ process failures \emph{at each step $s$}. 

Similarly with \emph{Redundant TSQR}, the robustness of the algorithm
increases as the need for robustness increases (see section
\ref{sec:ftca:rtsqr:rob}). The big difference with Redundant TSQR in
terms of robustness is that Self-Healing TSQR replaces the failed
processes. Therefore, this redundancy rate gives the number of failed
processes that can be accepted \emph{at each step}. For instance, 1
process can fail at step 1 ; it will be respawned and 3 additional
processes can fail at step 2. As a consequence, the total number of
failures this algorithm can tolerate is $\sum^{p}_{k=1} 2^k$. Besides,
at each step $s$ it can tolerate  $2^s - 1$ process failures.

\subsubsection{Behavior upon failures}
\label{sec:ftca:shtsqr:failure}

When a process fails, the process that was supposed to communicate
with it detects the failure and spawns a new process. The new process
obtains the redundant data from one of the processes that hold the
same data as the failed process. Then the computation continues
normally. 

\begin{figure}
  \resizebox{\linewidth}{!}{
\begin{tikzpicture}
\huge


\foreach \y in {0, 1, 2, 3}{
  \path[draw,thick] ( 0, -4*\y ) rectangle ( 1.8, -4*\y-3.8 );
    \node at ( -1, -4*\y-.8 ) {$\mathbf{P_\y}$ };
    \node at ( 1, -4*\y-.8 ) {$A_\y$ };
}


\foreach \y in {0, 1, 2, 3}{
  \path[draw,thick] ( 3.5, -4*\y ) rectangle ( 5.3, -4*\y-3.8 );
    \node at ( 4.7, -4*\y-.6 ) {$R_\y$ };
    \node at ( 4, -4*\y-2.5 ) {$V_\y$ };
    \draw [ thick] ( 3.5, -4*\y ) -- ( 5.3, -4*\y-1.8 );
    \draw [ ->, thick ] (2, -4*\y-1) -- (3.3, -4*\y-1 );
}
  \node at ( 2.7, 1.5 ) {\textit{QR} };


  \foreach \y in {0, 2}{
  \path[draw,thick] ( 7, -4*\y ) rectangle ( 8.8, -4*\y-3.6 );
    \node at ( 8.2, -4*\y-.6 ) {$R_\y$ };
    \pgfmathsetmacro\z{\y + 1};
    \node at ( 8.2, -4*\y-2.4 ) {$R_{\pgfmathprintnumber{\z}}$ };
   \draw [thick] ( 7, -4*\y-1.8 ) -- ( 8.8, -4*\y-1.8 );
  \draw [ thick] ( 7, -4*\y ) -- ( 8.8, -4*\y-1.8 );
    \draw [thick] ( 7, -4*\y-1.8 ) -- ( 8.8, -4*\y-3.6 );
     \draw [  ->, thick ] (5.5, -4*\y-5) -- (6.8, -4*\y-2.8 );
   \draw [  ->, thick ] (5.5, -4*\y-1) -- (6.8, -4*\y-1 );
}


  \foreach \y in {1, 3}{
  \path[draw,thick, dashed] ( 7, -4*\y ) rectangle ( 8.8, -4*\y-3.6 );
    \pgfmathsetmacro\z{\y-1};
    \node at ( 8.2, -4*\y-.6 ) {$R_{\pgfmathprintnumber{\z}}$ };
    \pgfmathsetmacro\z{\y};
    \node at ( 8.2, -4*\y-2.4 ) {$R_{\pgfmathprintnumber{\z}}$ };
   \draw [thick, dashed] ( 7, -4*\y-1.8 ) -- ( 8.8, -4*\y-1.8 );
  \draw [ thick, dashed] ( 7, -4*\y ) -- ( 8.8, -4*\y-1.8 );
    \draw [thick, dashed] ( 7, -4*\y-1.8 ) -- ( 8.8, -4*\y-3.6 );
     \draw [  ->, thick, dashed ] (5.5, -4*\y+3) -- (6.8, -4*\y-.8 );
   \draw [  ->, thick, dashed ] (5.5, -4*\y-1) -- (6.8, -4*\y-3 );
}

  \node at ( 6.2, 1.5 ) {\textit{S/R} };

  
    \node at ( 10.1, -9.2 )  {\Cross};
  \node at ( 10.1, -10.3 )  {CRASH};


  \path[draw,thick] ( 10.5, 0 ) rectangle ( 12.3, -3.6 );
    \node at ( 11.7, -.6 ) {$R_0'$ };
    \node at ( 11, -2.5 ) {${V_0}'$ };
    \draw [ thick] ( 10.5, 0 ) -- ( 12.3, -1.8 );
  \draw [  ->, thick ] (9, -1) -- (10.3, -1 );

  \foreach \y in {1, 3}{
  \path[draw,thick, dashed] ( 10.5, -4*\y ) rectangle ( 12.3, -4*\y-3.6 );
     \pgfmathsetmacro\z{\y-1};
   \node at ( 11.7, -4*\y-.6 ) {$R_{\pgfmathprintnumber{\z}}'$ };
    \node at ( 11, -4*\y-2.5 ) {${V_{\pgfmathprintnumber{\z}}}'$ };
    \draw [ thick, dashed] ( 10.5, -4*\y ) -- ( 12.3, -4*\y-1.8 );
  \draw [  ->, thick, dashed ] ( 9, -4*\y-1) -- (10.3, -4*\y-1 );
  }

  \node at ( 11, 1.5 ) {\textit{QR} };


   \node [rotate=90] at ( 11.8, -9.8 ) {respawn};
  \draw [  ->, thick] ( 12.5, -13) -- (13, -11.8 );
   \node [rotate = 65] at ( 13.4, -12.8 ) {copy};

  \path[draw,thick] ( 12.5, -8 ) rectangle ( 14.3, -11.6 );
    \node at ( 13.7, -8.6 ) {$R_2'$ };
    \node at ( 13, -10.5 ) {${V_2}'$ };
    \draw [ thick] ( 12.5, -8 ) -- ( 14.3, -9.8 );


  \path[draw,thick] ( 16.5, 0 ) rectangle ( 18.3, -3.6 );
  \draw [ thick] ( 16.5,0 ) -- ( 18.3, -1.8 );
    \draw [ thick] ( 16.5, -1.8 ) -- ( 18.3, -3.6 );
    \draw [thick] ( 16.5, -1.8 ) -- ( 18.3, -1.8 );
    \node at ( 17.7, -.6 ) {$R_0'$ };
    \node at ( 17.7, -2.4 ) {$R_2'$ };
  \draw [  ->, thick ] (13, -1) -- (16.3, -1 );
  \draw [  ->, thick ] (14.5, -9) -- (16.3, -3 );


  \path[draw,thick,loosely dashed] ( 16.5, -8 ) rectangle ( 18.3, -11.6 );
  \draw [ thick, loosely dashed] ( 16.5,-8 ) -- ( 18.3, -9.8 );
    \draw [ thick, loosely dashed] ( 16.5, -9.8 ) -- ( 18.3, -11.6 );
    \draw [thick, loosely dashed] ( 16.5, -9.8 ) -- ( 18.3, -9.8 );
    \node at ( 17.7, -8.6 ) {$R_0'$ };
    \node at ( 17.7, -10.4 ) {$R_2'$ };
  \draw [  ->, thick,loosely  dashed ] (13, -1) -- (16.3, -9 );
 \draw [  ->, thick,loosely  dashed ] (14.5, -9) -- (16.3, -11 );


  \path[draw,thick,  dashed ] ( 16.5, -4 ) rectangle ( 18.3, -7.6 );
  \draw [ thick,  dashed] ( 16.5,-4 ) -- ( 18.3, -5.8 );
    \draw [ thick,  dashed] ( 16.5, -5.8 ) -- ( 18.3, -7.6 );
    \draw [thick,  dashed] ( 16.5, -5.8 ) -- ( 18.3, -5.8 );
    \node at ( 17.7, -4.6 ) {$R_0'$ };
    \node at ( 17.7, -6.4 ) {$R_2'$ };
  \draw [  ->, thick, dashed ] (13, -5) -- (16.3, -5 );
 \draw [  ->, thick, dashed ] (14, -13) -- (16.3, -7 );

  \path[draw,thick,  dashed ] ( 16.5, -12 ) rectangle ( 18.3, -15.6 );
  \draw [ thick,  dashed] ( 16.5,-12 ) -- ( 18.3, -13.8 );
    \draw [ thick,  dashed] ( 16.5, -13.8 ) -- ( 18.3, -15.6 );
    \draw [thick,  dashed] ( 16.5, -13.8 ) -- ( 18.3, -13.8 );
    \node at ( 17.7, -12.6 ) {$R_0'$ };
    \node at ( 17.7, -14.4 ) {$R_2'$ };
  \draw [  ->, thick, dashed ] (13, -5) -- (16.3, -13 );
 \draw [  ->, thick, dashed ] (14, -13) -- (16.3, -15 );

  \node at ( 15.5, 1.5 ) {\textit{S/R} };


  \path[draw,thick] ( 20, 0 ) rectangle ( 21.8, -3.6 );
    \node at ( 21.2, -.6 ) {$R$ };
    \node at ( 20.5, -2.5 ) {$V$ };
  \draw [ ->, thick ] (18.5, -1) -- (19.8, -1 );
  \draw [ thick] ( 20,0 ) -- ( 21.8, -1.8 );

  \path[draw,thick, dashed] ( 20, -4 ) rectangle ( 21.8, -7.6 );
    \node at ( 21.2, -4.6 ) {$R$ };
    \node at ( 20.5, -6.5 ) {$V$ };
  \draw [ ->, thick,  dashed ] (18.5, -5) -- (19.8, -5 );
  \draw [ thick,  dashed] ( 20, -4 ) -- ( 21.8, -5.8 );

  \path[draw,thick, loosely dashed] ( 20, -8 ) rectangle ( 21.8, -11.6 );
    \node at ( 21.2, -8.6 ) {$R$ };
    \node at ( 20.5, -10.5 ) {$V$ };
  \draw [ ->, thick, loosely dashed ] (18.5, -9) -- (19.8, -9 );
  \draw [ thick, loosely dashed] ( 20, -8 ) -- ( 21.8, -9.8 );

  \path[draw,thick, dashed] ( 20, -12 ) rectangle ( 21.8, -15.6 );
    \node at ( 21.2, -12.6 ) {$R$ };
    \node at ( 20.5, -14.5 ) {$V$ };
  \draw [ ->, thick, dashed ] (18.5, -13) -- (19.8, -13 );
  \draw [ thick, dashed] ( 20, -12 ) -- ( 21.8, -13.8 );

  \node at ( 19, 1.5 ) {\textit{QR} };
\end{tikzpicture}
  }
\captionof{figure}{\label{fig:shtsqr:crash}Self-Healing TSQR on 4
  processes with one process failure.} 
\end{figure}
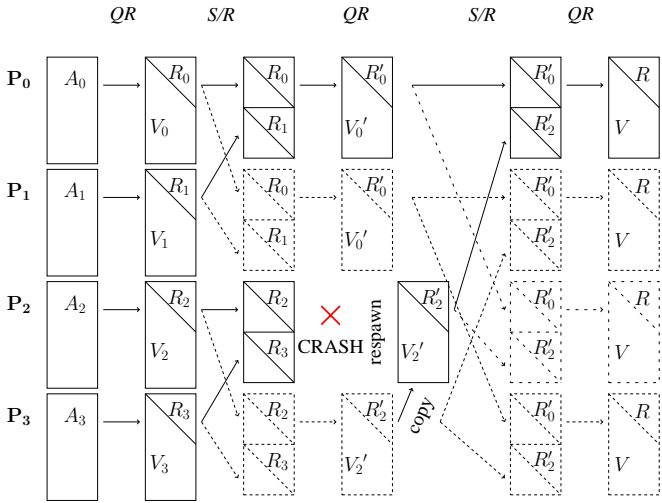

\footnotesize
\bibliographystyle{plain}
\bibliography{ft}

\begin{thebibliography}{10}

\bibitem{ACDHL10}
Emmanuel Agullo, Camille Coti, Jack Dongarra, Thomas Herault, and Julien
  Langou.
\newblock Qr factorization of tall and skinny matrices in a grid computing
  environment.
\newblock In {\em 24th IEEE International Parallel \& Distributed Processing
  Symposium (IPDPS'10)}, Atlanta, Ga, April 2010.

\bibitem{BBHHBD13}
Wesley Bland, Aurelien Bouteiller, Thomas H{\'{e}}rault, Joshua Hursey, George
  Bosilca, and Jack~J. Dongarra.
\newblock An evaluation of user-level failure mitigation support in {MPI}.
\newblock {\em Computing}, 95(12):1171--1184, 2013.

\bibitem{BDDL09}
George Bosilca, Remi Delmas, Jack Dongarra, and Julien Langou.
\newblock Algorithm-based fault tolerance applied to high performance
  computing.
\newblock {\em J. Parallel Distrib. Comput.}, 69(4):410--416, 2009.

\bibitem{BLKC04}
Aur\'elien Bouteiller, Pierre Lemarinier, G\'eraud Krawezik, and Franck
  Cappello.
\newblock Coordinated checkpoint versus message log for fault tolerant {MPI}.
\newblock {\em International Journal of High Performance Computing and
  Networking (IJHPCN)}, (3), 2004.

\bibitem{fgcs08}
Darius Buntinas, Camille Coti, Thomas Herault, Pierre Lemarinier, Laurence
  Pilard, Ala Rezmerita, Eric Rodriguez, and Franck Cappello.
\newblock Blocking vs. non-blocking coordinated checkpointing for large-scale
  fault tolerant {MPI}.
\newblock {\em Future Generation Computer Systems}, 24 (1):73--84, 2008.
\newblock Digital Object Identifier:
  http://dx.doi.org/10.1016/j.future.2007.02.002.

\bibitem{CFGLABD05}
Zizhong Chen, Graham~E Fagg, Edgar Gabriel, Julien Langou, Thara Angskun,
  George Bosilca, and Jack Dongarra.
\newblock Fault tolerant high performance computing by a coding approach.
\newblock In {\em Proceedings of the tenth ACM SIGPLAN symposium on Principles
  and practice of parallel programming}, pages 213--223. ACM, 2005.

\bibitem{CAQR}
James Demmel, Laura Grigori, Mark Hoemmen, and Julien Langou.
\newblock Communication-avoiding parallel and sequential {QR} factorizations.
\newblock {\em CoRR}, abs/0806.2159, 2008.

\bibitem{DGG10}
Simplice Donfack, Laura Grigori, and Alok~Kumar Gupta.
\newblock Adapting communication-avoiding lu and qr factorizations to multicore
  architectures.
\newblock In {\em Parallel \& Distributed Processing (IPDPS), 2010 IEEE
  International Symposium on}, pages 1--10. IEEE, 2010.

\bibitem{DBBHD12}
Peng Du, Aurelien Bouteiller, George Bosilca, Thomas Herault, and Jack
  Dongarra.
\newblock Algorithm-based fault tolerance for dense matrix factorizations.
\newblock {\em ACM SIGPLAN Notices}, 47(8):225--234, 2012.

\bibitem{FTMPI}
Graham~E. Fagg and Jack Dongarra.
\newblock {FT}-{MPI}: Fault tolerant {MPI}, supporting dynamic applications in
  a dynamic world.
\newblock In Jack Dongarra, P{\'e}ter Kacsuk, and Norbert Podhorszki, editors,
  {\em Recent Advances in Parallel Virtual Machine and Message Passing
  Interface, 7th European {PVM}/{MPI} Users' Group Meeting, Balatonf{\"u}red,
  Hungary, September 2000, Proceedings}, volume 1908 of {\em Lecture Notes in
  Computer Science}, pages 346--353. Springer, 2000.

\bibitem{FGBACPLD04}
Graham~E Fagg, Edgar Gabriel, George Bosilca, Thara Angskun, Zizhong Chen,
  Jelena Pjesivac-Grbovic, Kevin London, and Jack~J Dongarra.
\newblock Extending the mpi specification for process fault tolerance on high
  performance computing systems.
\newblock In {\em Proceedings of the International Supercomputer Conference
  (ICS)}, 2004.

\bibitem{mpi3}
Message Passing~Interface Forum.
\newblock {MPI}: {A} message-passing interface standard, version 3.0.
\newblock Technical report, 2012.

\bibitem{mpi-3.1}
Message Passing~Interface Forum.
\newblock {MPI: A Message-Passing Interface Standard Version 3.1}, 09 2012.

\bibitem{HLAD09}
B.~Hadri, H.~Ltaief, E.~Agullo, and J.~Dongarra.
\newblock Tall and skinny qr matrix factorization using tile algorithms on
  multicore architectures.
\newblock Technical report, Innovative Computing Laboratory, University of
  Tennessee, September 2009.

\bibitem{HGBBPS11}
Joshua Hursey, Richard~L. Graham, Greg Bronevetsky, Darius Buntinas, Howard
  Pritchard, and David~G. Solt.
\newblock Run-through stabilization: An {MPI} proposal for process fault
  tolerance.
\newblock In Yiannis Cotronis, Anthony Danalis, Dimitrios~S. Nikolopoulos, and
  Jack Dongarra, editors, {\em Recent Advances in the Message Passing Interface
  - 18th European {MPI} Users' Group Meeting, EuroMPI 2011, Santorini, Greece,
  September 18-21, 2011. Proceedings}, volume 6960 of {\em Lecture Notes in
  Computer Science}, pages 329--332. Springer, 2011.

\bibitem{L10}
Julien Langou.
\newblock Computing the r of the qr factorization of tall and skinny matrices
  using mpi\_reduce.
\newblock {\em arXiv preprint arXiv:1002.4250}, 2010.

\bibitem{PLP98}
James~S. Plank, Kai Li, and Michael~A. Puening.
\newblock Diskless checkpointing.
\newblock {\em IEEE Trans. Parallel Distrib. Syst.}, 9(10):972--986, October
  1998.

\bibitem{Reed2006293}
Daniel~A. Reed, Charng da~Lu, and Celso~L. Mendes.
\newblock Reliability challenges in large systems.
\newblock {\em Future Generation Computer Systems}, 22(3):293 -- 302, 2006.

\end{thebibliography}

\end{document}